\documentclass[fleqn,twoside]{article}
\usepackage{espcrc2}

\usepackage{graphicx}
\usepackage{subfigure}
\usepackage[figuresright]{rotating}

\newcommand{\AmS}{{\protect\the\textfont2
  A\kern-.1667em\lower.5ex\hbox{M}\kern-.125emS}}

\title{A New Parameterization of the Nucleon Elastic Form Factors}

\author{R. Bradford,\address[uofr]{Department of Physics and Astronomy, University of Rochester, Rochester NY  14627-0171, USA}%
  A. Bodek,\addressmark[uofr]
  H. Budd,\addressmark[uofr]
  and J. Arrington\address{Argonne National Laboratory 
        Argonne, IL 60439, USA}}

\begin{document}

\begin{abstract}
The nucleon elastic form factors are generally interpreted as a mapping of the charge and magnetic current distributions of the proton and neutron.  New high $Q^2$ measurements have opened up fundamental questions about $G_{ep}$ that remain to be answered.  This talk will summarize current developments surrounding the nucleon form factors and explain why they are important to neutrino physicists.  New parameterizations of the nucleon form factors, suitable for use by neutrino physicists,  will be introduced and discussed.
\vspace{1pc}
\end{abstract}

\maketitle

\section{Introduction}
While the nucleon elastic form factors have been measured for 50 years in $e^- N$ scattering, recent measurements from Jefferson Lab have shown unexpected structure in the ratio of $\frac{\mu _p G_{ep}}{G_{mp}}$.  Understanding the new measurements has been a major focus of the Jlab community.  As the elastic nucleon form factors are input for neutrino simulations packages, it is important for neutrino physicists to understand the nucleon form factors and the current controversy.

This talk will begin by presenting an overview of the nucleon form factors in Section \ref{oview}.  Section \ref{measurements} will discuss two techniques for measuring the form factors and briefly discuss the new Jefferson Lab measurements.  Section \ref{neutrino} motivates the role of the nucleon form factors in neutrino physics.  The talk ends by presenting a new parametrization of the form factors in Sections \ref{parameterization} and \ref{discussion}.

\section{Overview}
\label{oview}
In the single-photon exchange approximation, the nucleon elastic form factors arise in the elastic electron-nucleon scattering cross section according to:
\begin{equation}
\frac{d \sigma}{d \Omega}= \frac{\alpha^2 E'_{e} \cos \left( \frac{\theta _e}{2} \right)}{4 E^3_{e} \sin ^4\left( \frac{\theta _e}{2} \right)} \left[ G_{eN}^2 + \frac{\tau}{\varepsilon}G^2_{mN} \right] \left(\frac{1}{1+\tau}\right)
\label{f1}b
\end{equation}
where $E$ is the incident electron energy, $E'$ is the scattered electron energy, $\theta _e$ is the electron scattering angle, and $\tau=\frac{Q^2}{4M^2}$ (with $M$ being the nucleon mass).  $\varepsilon = \left[ 1+ 2\left( 1+\tau \right) \tan ^2 \frac{\theta _e}{2} \right] ^{-1}$ is the polarization of the exchange photon mediating the interaction.  $G_{eN}$ is the nucleon electric form factor, and $G_{mN}$ is the magnetic form factor.  While unique form factors exist for both nucleons, this paper will use $G_{eN}$ and $G_{mN}$ when making statements that may be applied to both the proton and neutron.

The form factors account for the effects of the spatial size of the nucleons on the elastic cross section.  In Equation (\ref{f1}), the overall coefficient of $\frac{\alpha^2 E'_{e} \cos \left( \frac{\theta _e}{2} \right)}{4 E^3_{e} \sin ^4\left( \frac{\theta _e}{2} \right)}$ is known as the Mott cross section, which is the standard cross section for elastic scattering of point-like particles.  The Mott cross section was originally thought to explain elastic electron-nucleon scattering.  Rosenbluth developed a cross section formula in 1950 that introduced a ``form factor'' to account for the case that the nucleon may not be point-like particle.  Rosenbluth's formula came into wide use after the Mott cross section formula failed to explain early measurements from Hofstadter and McAllister \cite{hofstadter}.

While there are various parameterizations of the elastic nucleon differential cross section, Equation (\ref{f1}) employs the Sachs form factors. In the non-relativistic limit, these may be interpreted as the Fourier transform of the nucleon charge and current distributions.

\section{Experimental Measurements}
\label{measurements}
Over the years, two different techniques for measuring the elastic form factors have been developed.  The first, and oldest technique, is to perform a Rosenbluth separation, while the ``newer'' technique involves using recoil polarization measurements to extract the form factor ratio.  The next two sections will discuss each type of measurement.

\subsection{Early measurements:  Rosenbluth Separation}
The earliest form factor measurements were made in the 1950's using a technique known as Rosenbluth separation.  Rosenbluth separation takes advantage of Equation (\ref{f1})'s linear dependence on $\varepsilon$.  The idea was fairly straightforward:  The elastic cross section was measured at various values of $\varepsilon$ by holding $Q^2$ constant while varying $\theta _e$.  A line was fit to the resulting cross section, and the fit parameters yielded the various form factors - the intercept gave a measurement of $\tau G_{mN}^2$, while the slope yielded $G _{eN}^2$.

Early measurements of the form factors appeared to be fit well by a parameterization known as the dipole form factor:
\begin{equation}
G_d=\left( 1+\frac{Q^2}{\Lambda ^2}\right)^{-2}
\end{equation}
where $\Lambda^2$=0.71 GeV$^2$.  Three of the four form factors, $G_{ep}$, $\frac{G_{mp}}{\mu _p}$, and $\frac{G _{mn}}{\mu _n}$ were well modeled by this parameterization.

Despite the early success of the Rosenbluth separation, the method did have some weaknesses.  Because the method involved first measuring the elastic cross section, it was susceptible to the systematic errors that are inherent in cross section measurements.  In addition, the method could only produce precise measurement of $G _{eN}$ below $Q^2=1 GeV^2$.  At higher $Q^2$, the $G _{eN}$ term is damped by a factor of $\frac{1}{\tau}$, as seen in Equation (\ref{f1}).  The cross section, then, becomes dominated by $G _{mN}$ above $Q^2= 1 GeV ^2$.  Finally, the whole formalism rests on the assumption of single photon exchange.  While the measured cross sections are correct for terms beyond one photon exchange, the two photon exchange corrections to the form factors are not well understood.  Significant two photon contributions to the cross section could undermine the validity of the formalism.
 
\subsection{Polarization Measurements}
A second method of measuring the elastic form factors involves scattering polarized electrons from the nucleon, and then measuring orthogonal components of the nucleon's recoil polarization.  The ratio of the polarization components is related to the ratio of the electric and magnetic form factors by
\begin{equation}
\frac{G_{eN}}{G_{mN}} = -\frac{P_t}{P_l}\frac{\left(E_e + E'_e\right)}{2M}\tan \left(\frac{\theta _e}{2}\right), 
\end{equation}
where $P_l$ and $P_t$ are longitudinal and transverse (with respect to the nucleon momentum) components of the nucleon's polarization.  

\begin{figure}
\begin{center}
\includegraphics[width=2.5in]{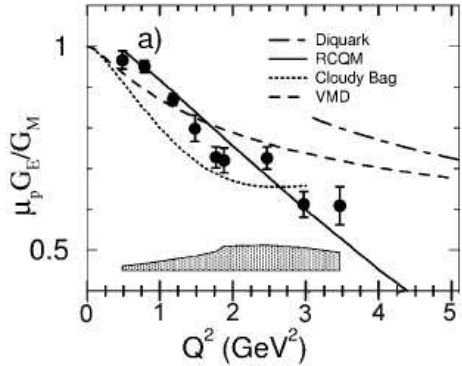}
\caption{$\frac{\mu _p G_{ep}}{G_{mp}}$ measured in Hall A at Jefferson Lab \cite{jlab1}.  Based on previous measurements, we expected $\frac{\mu _p G_{ep}}{G_{mp}}$ to be flat and ~1.  However, the new results here based on recoil polarization measurements drop linearly in $Q^2$.  The curves are for various model calculations, references to which can be found in \cite{jlab1}.  The shaded bar towards the bottom of the plot represents expected systematic errors.}
\label{jfig1}
\end{center}
\end{figure}

Because the measurement involves a ratio of polarization components, many of the possible systematic errors traditionally associated with the use of a polarimeter divide out of the measurement.  This method is viewed as being systematically more robust than measurements made with a Rosenbluth separation.

\subsection{Recent Jlab Measurements}
While the recoil polarization measurements have been made since the 1970's, it was not until recently that the two methods were shown to disagree.  Figure \ref{jfig1} shows data published in 2000 that was taken in Hall A at Jefferson Lab \cite{jlab1}.  The figure shows the ratio $\frac{\mu _p G_{ep}}{G_{mp}}$.  Based on experience from earlier measurements employing the Rosenbluth Separation technique, we expected this ratio to be approximately one as a function of $Q^2$.  However, the newer data drop off linearly with $Q^2$.  

This was a great surprise to the nuclear physics community.  The result has spawned much dialog, scrutiny of older datasets, and a number of additional experiments attempting to verify, refute, or explain the discrepancy.  Further recoil polarization results from Jlab using the same experimental setup show that the discrepancy persists at higher values of $Q^2$ \cite{jlab2,jlab3}, while re-analysis of older datasets show that the Rosenbluth measurements are self-consistent \cite{arring1}.  Results from a ``Super-Rosenbluth'' experiment, which attempted a precise Rosenbluth separation having systematic errors comparable to the recoil polarization experiments, are also consistent with older measurements based on the Rosenbluth separation method \cite{superosenbluth}. 

While the source of the discrepancy remains an open question, current efforts focus on possible two-photon contributions to the elastic $ep$ cross section \cite{arring2,arring3,blunden,guichon}.  The extent of this effect is currently under investigation.  If this mechanism proves to explain the discrepancy, then the resulting errors will be more pronounced in the case of the Rosenbluth form factors, making form factors based on the polarization transfer technique preferred.  

\section{Neutrino Physics}
\label{neutrino}
The elastic nucleon form factors have direct bearing on neutrino physics.  The vector part of the neutrino cross section can be expressed in terms of $G^V _E$ and $G^V _M$, the vector electric and magnetic form factors.  Through the conserved vector current hypothesis, these form factors may then be related to the elastic nucleon form factors measured in elastic $eN$ scattering as shown by:
\begin{equation}
G^V _E \left( Q^2 \right) = G _{ep}\left( Q^2 \right) - G _{en} \left( Q^2 \right)
\end{equation}
and 
\begin{equation}
G^V _M \left( Q^2 \right) = G _{mp}\left( Q^2 \right) - G _{mn} \left( Q^2 \right).
\end{equation}

Current neutrino simulation programs use the elastic nucleon form factors to parameterize the vector part of the elastic $\nu A$ cross section.  In light of the recent controversy, it is important for neutrino physicists to understand the state of nucleon form factor measurements and realize that there are open questions that are currently being investigated.  Attention must be paid to the field as it develops over the next fews years, and parameterizations selected for used in simulations must be chosen carefully.  Ideally, one should employ a parameterization with reasonable constraints at both low- and high-$Q^2$.

\begin{sidewaystable}
\caption{Fit parameters, given $G_{en}>0$ and $\frac{d}{u}$=0.2.  The $a_0$ parameter was used to ensure the correct low $Q^2$ limit and was not varied during the fits.}
\label{t1}
\begin{tabular*}{\textheight}{@{\extracolsep{\fill}}lllllllllllll}
\hline
Observable & $a_0$ & $a_1$ & $a_2$ & $b_1$ & $b_2$ & $b_3$ & $b_4$\\
\hline
$G_{Ep}$ & $1$ & $-0.0578 \pm .166$ & & $11.1 \pm 0.217$ & $13.6 \pm 1.39$ & $33.0 \pm 8.95$\\
$G_{Mp}$ & $1$ & $0.150 \pm 0.0312$ & & $11.1 \pm 0.103$ & $19.6 \pm 0.281$ & $7.54 \pm 0.967$\\
$G_{En}$ & $0$ & $1.25 \pm 0.368$ & $ 1.30 \pm 1.99 $ & $ -9.86 \pm 6.46 $ & $ 305 \pm 28.6$ & $ -758 \pm 77.5$ & $ 802 \pm 156$\\
$G_{Mn}$ & $1$ &$ 1.81 \pm 0.402$ & & $ 14.1 \pm 0.597$ & $ 20.7 \pm 2.55 $ & $68.7 \pm 14.1$\\
\hline
\end{tabular*}
\end{sidewaystable}

\section{New Parameterizations}
\label{parameterization}
The recent controversy has lead physicists to question the validity of the $G_d$ parameterization.  Many new parameterizations have been developed \cite{bba,kelly} based on fits to experimental data.  

We have developed a new parameterization which builds on earlier work from our group \cite{bba} and efforts by Kelly \cite{kelly}.  The parameterization was developed by fitting a single functional form for all four elastic form factors.  Datasets used in the fitting were similar to those used by Kelly \cite{kelly}, although we did not include measurements of mean nucleon radius in our fits.  The data emphasized measurements based on the polarization transfer technique and excluded Rosenbluth measurements of $G_{ep}$ above $Q^2>1 GeV^2$.  For this analysis, we have excluded the data requiring large two photon exchange corrections in the form factor extraction.  A more complete analysis, currently underway, would correct both the cross section and polarization results for two photon exchange, but is beyond the scope of this work.

The functional form is given by 
\begin{equation}
G \left( Q^2 \right) = \frac{\sum_{k=0}^n a_k \tau ^k}{1+\sum_{k=1}{b_k \tau^k}}.\end{equation}  
While this form has been used by other parameterizations in the past \cite{kelly}, this is the first time that this particular form has been employed for all four form factors.  We will refer to our parameterizations as the ``BBBA05'' form factors throughout the rest of this talk.

An additional feature of our new parameterizations is the implementation of two constraints applied in the fitting.  The first constraint comes from local duality.  $R$ is defined as the ratio of form factors.  In the elastic limit, $R$ takes the form 
\begin{equation}
R_N\left( x=1; Q^2 \right) = \frac{4M^2}{Q^2}\left( \frac{G^2 _{eN}}{G^2_{mN}}\right)
\end{equation}
As $Q^2 \rightarrow \infty$, $R_n=R_p$, so our first constraint takes the form
\begin{equation}
\left(\frac{G_{en}}{G_{mn}}\right)^2=\left(\frac{G_{ep}}{G_{mp}}\right)^2
\end{equation}

\begin{figure*}
\begin{center}
\includegraphics[width=4in, height=2.5in]{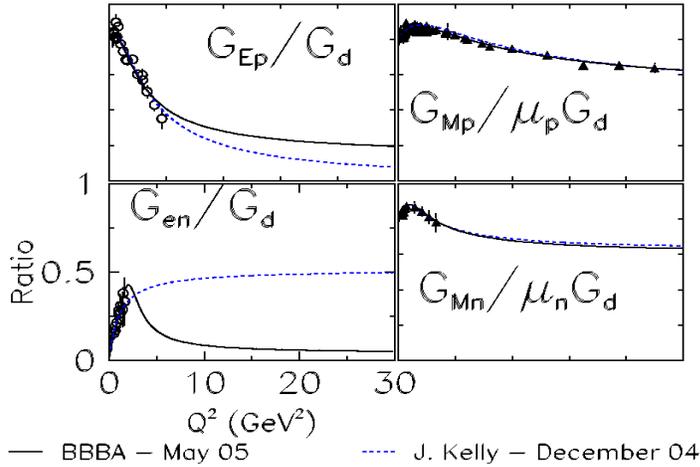}
\caption[Form factor ratios to $G_d$]{The solid black line shows the ratio of the BBBA05 form factors to $G_d$, and the dashed blue line is the ratio of the Kelly form factors to $G_d$.  The differences in the two parameterizations  for $\frac{G_{ep}}{G_d}$ and $\frac{G_{en}}{G_d}$ are due to the constraints applied to the BBBA05 form factors.  All figures have a y-axis ranging from $Q^2=0 GeV^2$ to $Q^2=30 GeV^2$.  In the lower limit ($Q^2=0GeV^2$), all ratios approach unity, except for $G_{en}$, which approaches zero.}
\label{fits1}
\end{center}
\end{figure*}

A second constraint is based on QCD sum rules and a further application of duality.  In the elastic limit, we can express the $\frac{F_{2n}}{F_{2p}}$ ratio as
\begin{equation}
\left( \frac{F_{2n}}{F_{2p}} \right)^2 = \left( \frac{G_{mn}}{G_{mp}} \right)^2
\label{step1}
\end{equation}
In the $\nu \rightarrow \infty$ and $Q^2 \rightarrow \infty$ and fixed $x$ limits, the $F_2$ form factor becomes a simple quark-counting exercise
\begin{equation}
F_2=x \sum_{i}e_i^2 f_i \left(x \right).
\label{step2}
\end{equation}
Inserting (\ref{step2}) into (\ref{step1}), we arrive at our second constraint:
\begin{equation}
\left( \frac{G_{mn}}{G_{mp}} \right)^2 = \frac{1+4\frac{d}{u}}{4+\frac{d}{u}}
\label{constraint2}
\end{equation}
The value of $\frac{d}{u}$ is somewhat subjective.  We actually ran three sets of fits, each with a different value of $\frac{d}{u}=$0, 0.2, or 0.5.  Our preferred value was $\frac{d}{u}=$0.2.

The above constraints were implemented by scaling the high $Q^2$ data-points of $G_{mp}$ and then adding these scaled points to the datasets for $G_{en}$ and $G_{mn}$ during the fits.  The ``constraint data'' are not shown in the figures of this paper.  While we initially tried to apply the constraints explicitly to the fit parameters that determined the high $Q^2$ behavior of the less well-measured form factors, the convergence of the constraints was very slow.  We were looking for something to converge around $Q^2=30 GeV^2$.  Using these additional data-points satisfied this criteria.  Errors on the ``constraint data'' were inflated to keep these additional points from wielding too strong an influence on the fits.

Because the above constraints are all on squares of form factors, one may argue about different sign conventions that one could use in the application of the constraints, particularly at high $Q^2$, where many of the form factors are poorly measured.  At large $Q^2$ values, where $G_{ep}$ and $G_{en}$ might change sign, their contributions to the neutrino cross section are extremely small \cite{bba03}.  We ran fits with both signs for $G_{en}$ and preferred the positive $G_{en}$ in the end.  $G_{en}<0$ yielded odd oscillatory behavior in the constraint at the sign change.

The plots that will be shown in this paper are based on $G_{en}>0$, and with $\frac{d}{u}=0.2$.  The fit parameters are shown in Table \ref{t1}.

\section{Plots and discussion of new parameterization}
\label{discussion}
Plots of the new parameterizations are shown in Figure \ref{fits1}.  As shown in the figure, the BBBA05 parameterizations of $G_{Mp}$ and $G_{Mn}$ are close to the Kelly form factors.  However, the new functional form and added high $Q^2$ constraints cause the BBBA05 parameterization of $G_{En}$ to die off much more quickly at high $Q^2$ than does the Kelly parameterization.  

\begin{figure}
\begin{center}
\includegraphics[width=2.55in]{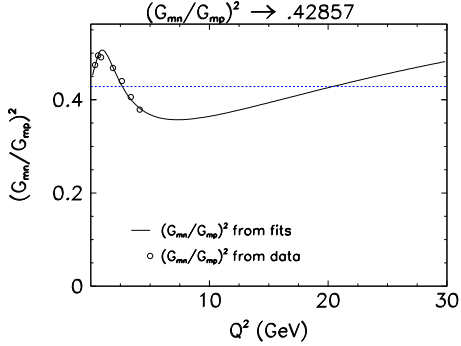}
\caption[$\left( \frac{G_{mn}}{G_{mp}} \right) ^2$ at high $Q^2$]{The effect of constraining $\left( \frac{G_{mn}}{G_{mp}} \right) ^2$ at high $Q^2$ is demonstrated here.  $\frac{G_{mn}}{G_{mp}}$ for the BBBA05 parameterization (solid black line) intersects the asymptotic value (blue dashed line) at a single point around $Q^2=20GeV^2$.  Data points are average ratios of available data across bins 250 $MeV^2$-wide in $Q^2$ where data exist for the appropriate form factors.}
\label{c1}
\end{center}
\end{figure}

\begin{figure}
\begin{center}
\includegraphics[width=2.55in]{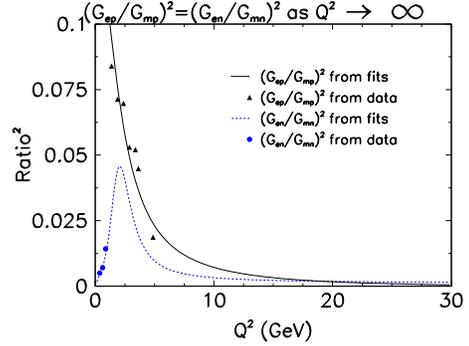}
\caption[$\left( \frac{G_{ep}}{G_{mp}} \right) ^2$ and $\left( \frac{G_{en}}{G_{mn}} \right)$ at high $Q^2$]{The two curves represent  $\left( \frac{G_{ep}}{G_{mp}} \right) ^2$ (solid black) and $\left( \frac{G_{en}}{G_{mn}} \right)$ (dashed blue) at high $Q^2$. Data points are average ratios of available data across bins 250 $MeV^2$-wide in $Q^2$ where data exist for the appropriate form factors.}
\label{c2}
\end{center}
\end{figure}

Plots demonstrating the behavior of the constraints are shown in Figures \ref{c1} and \ref{c2}.  The ratios here both appear to satisfy the constraint at one point around $Q^2=20 GeV^2$.  This behavior is a related to our implementation of the constraints as additional data-points (data from $G_{mp}$ scaled according to the constraints and added to the fit datasets for $G_{en}$ and $G_{mn}$).  Hence, we claim that our parameterization is valid up to approximately $Q^2=18 GeV^2$.

\section{Acknowledgments}
Work supported in part by the US Department of Energy, Office of Nuclear Physics, under contract W-31-109-ENG-38, and the DOE Office of High Energy Physics under grant DE-FG02-91ER40685.

\end{document}